\documentclass[
journal=jctcce,
manuscript=article,
layout= twocolumn
]{achemso}
\setkeys{acs}{maxauthors=10, etalmode=truncate}
\usepackage[fontsize=9pt]{fontsize}
\DeclareCaptionTextFormat{baseline}{\baselineskip 9.5pt #1}

\usepackage{xcolor}
\usepackage{mathtools}
\usepackage{upgreek}
\usepackage{amssymb}

\renewcommand{\vec}[1]{{\boldsymbol{#1}}}
\renewcommand{\matrix}[1]{{\boldsymbol{#1}}}

\renewcommand{\d}{\mathrm{d}}
\DeclareSymbolFont{rmlargesymbols}{OMX}{mdbch}{m}{n}
\DeclareMathSymbol{\rmintop}{\mathop}{rmlargesymbols}{82}
\newcommand{\upint}{\rmintop\nolimits}

\newcommand{\E}[1]{\left< #1 \right>}
\newcommand{\cE}[2]{\left< #1 \middle| #2 \right>}

\newcommand{\euler}{\mathrm{e}}
\newcommand{\kb}{\mathrm{k}_\mathrm{B}}
\newcommand{\itGamma}{{\it{\Gamma}}}
\newcommand{\itPi}{{\it{\Pi}}}
\newcommand{\us}{{\textmu}{s}}

\newcommand{\T}{\mathrm{T}}

\definecolor{myred}{HTML}{bc2a13}
\definecolor{mygreen}{HTML}{39aa26}

\newcommand{\SIFigAlaMass}       {S1}
\newcommand{\SIFigAlaConvergence}{S2}
\newcommand{\SIFigAlaReverse}    {S3}
\newcommand{\SIFigAlaPathratio}  {S4}
\newcommand{\SIFigLysConvergence}{S5}
\newcommand{\SIFigLysFriction}   {S6}

\author{Matthias Post}
\altaffiliation{present address: Max Planck Institute of Biophysics, Frankfurt on Main, Germany}

\author{Steffen Wolf}

\author{Gerhard Stock}
\email{stock@physik.uni-freiburg.de}

\affiliation{Biomolecular Dynamics, Institute of Physics, University of Freiburg, Freiburg 79104, Germany}

\title{Investigation of rare protein conformational transitions via
dissipation-corrected targeted molecular dynamics}

\begin{document}

%
%
%

\begin{abstract}
To sample rare events, dissipation-corrected targeted molecular
  dynamics (dcTMD) applies a constant velocity constraint along a
  one-dimensional reaction coordinate $s$, which drives an atomistic
  system from an initial state into a target state. Employing a
  cumulant approximation of Jarzynski's identity, the free energy
  $\Delta G (s)$ is calculated from the mean external work
  and dissipated work of the
  process. By calculating the friction coefficient $\itGamma (s)$ from  
  the dissipated work, in a second step the equilibrium dynamics of
  the process can be studied by propagating a Langevin equation.
  While so far dcTMD has been mostly applied to study the unbinding of
  protein-ligand complexes, here its applicability to rare
  conformational transitions within a protein and the prediction of their kinetics is investigated. As this
  typically requires the introduction of multiple collective
  variables $\{x_j\}= \vec{x}$, a theoretical framework is outlined
  to calculate the associated free energy $\Delta G (\vec{x})$ and
  friction $\matrix{\itGamma}(\vec{x})$ from dcTMD simulations along
  coordinate $s$. Adopting the $\alpha$-$\beta$ transition of alanine
  dipeptide as well as the open-closed transition of T4 lysozyme as representative examples, the
  virtues and shortcomings of dcTMD to predict protein conformational
  transitions and the related kinetics are studied.
\end{abstract}

\section{Introduction}
\label{sec:introduction}

In computational biophysics, molecular dynamics (MD) simulations provide a microscopic description of protein dynamics with a resolution in both space and time that is inaccessible for all currently existing experimental methods. MD simulations are therefore a valuable tool to complement experiments.\cite{Shaw10} However, two standing challenges of this computational approach are a) that due to the necessity to employ a femtosecond-range time step, processes on time scales longer than a millisecond cannot be accessed within reasonable real-world time\cite{Shaw21} and b) the full microscopic information of such simulations is not comprehensible for humans. 

Since only a subset of all degrees of freedom of a biological macromolecule are essential for the functionally relevant processes, they may instead be described by a small set of $n$ collective variables (CVs) $\vec{x} = (x_1,\dots,x_n)$.\cite{Sittel18} If a timescale separation between the slow motion along the CVs and the fast fluctuations of the remaining degrees of freedom exist, the dynamics may be modeled by a Langevin equation \cite{Zwanzig01} of the form
\begin{equation} \label{eq:langevin_equation}
  \matrix{M} \ddot{\vec{x}}  = - \frac{\partial G}
  {\partial \vec{x}} - \matrix{\itGamma}(\vec{x})
  \, \dot{\vec{x}}  + \vec{\xi}(t) \; ,
\end{equation}
where $\matrix{M}$ denotes the mass matrix, $G$ represents the free energy landscape resulting in a mean force, $\matrix{\itGamma}$ is a position-dependent Stokes friction coefficient, and $\vec{\xi}$ represents Gaussian white noise with the amplitude given by the fluctuation-dissipation theorem. If the fields $G(\vec{x})$ and $\matrix{\itGamma}(\vec{x})$ are known, the protein's essential dynamics can be simulated via Eq.~\eqref{eq:langevin_equation} with considerably less computational effort than of a fully atomistic MD simulation. Estimating these fields from equilibrium simulation\cite{Lange06b,Hegger09,Ayaz21}, on the other hand, is only possible if all relevant states are visited, at which point we wouldn't need a coarse-grained model from the outset.

To resolve this issue, enhanced sampling techniques\cite{Henin22} may be employed to
accelerate these rare transitions and thus estimate $G$ and $\matrix{\itGamma}$.
Popular examples that allow for the estimation of unbiased 
rates\cite{Wolf23a} are 
random acceleration MD\cite{Kokh18,NunesAlves21}, weighted ensemble\cite{Votapka17} and  milestoning\cite{Ojha23} approaches, infrequent metadynamics\cite{Tiwary13,Casasnovas17} or Gaussian-accelerated MD\cite{Miao20,Wang22a}. 
In this work, we focus on dissipation-corrected targeted MD (dcTMD)\cite{Wolf18}, which utilizes the idea of TMD\cite{Schlitter93A} by constraining the system along a suitable reaction coordinate $s$ with constant velocity $v$, i.e.,
\begin{equation} 
	s(t) = s_0 + v t ,
\end{equation}
in order to enforce the transition between two states of interest. By calculating the work $W$ from the applied constraint force $f$,
\begin{equation}
 W(s) = \upint_0^s \!\! \d s^\prime f(s^\prime) ,
 \label{eq:work}
\end{equation}
the free energy can be estimated from Jarzynski's
equation\cite{Jarzynski97} 
\begin{subequations}\label{eq:free_energy_reaction_coordinate}
 \begin{align} 
\Delta G(s) &= -\beta^{-1} \ln \cE{ \euler^{-\beta W}}{s} \label{eq:free_energy_reaction_coordinate_exp} \\
&\approx \cE{W}{s} - \frac{\beta}{2} \cE{ \delta W^2}{s} \; , \label{eq:free_energy_reaction_coordinate_cum}
 \end{align}
\end{subequations}
where $\cE{\cdot}{s}$ denotes an average over pulling simulations along coordinate $s$ starting from an equilibrium state at $s=s_0$, and $\beta = 1/(\kb T)$ is the inverse temperature. The cumulant approximation in Eq.~\eqref{eq:free_energy_reaction_coordinate_cum} is appropriate if the work distribution is close to a Gaussian with $\delta W = W -\E{W}$.

To run Langevin dynamics using Eq.~\eqref{eq:langevin_equation}, we
furthermore need the friction $\itGamma (s)$, which can be calculated
from the average dissipated energy $W_\text{diss} = W - \Delta G = \beta \left\langle \delta W^2 \right\rangle / 2$ via
\cite{Wolf18,Post22} 
\begin{subequations}
 \begin{align} 
\itGamma(s) &= \frac{1}{v}\frac{\d \cE{ W_\text{diss}}{s}}{\d s} \label{eq:friction_reaction_coordinate_exp}\\
&\approx \beta \upint_0^{t(s)} \!\!\! \d \tau \E{ \delta f(t) \delta f(\tau) } \; , \label{eq:friction_reaction_coordinate_cum}
 \end{align}
\end{subequations}
where $\delta f = f - \E{f}$ and the second line again invokes the cumulant approximation.

In previous work, we focused on the application of dcTMD on biomolecular systems where a one-dimensional description of the unbinding process by the protein-ligand distance appears quite natural. Recent examples for such systems are the unbinding of protein--ligand complexes such as trypsin--benzamidine\cite{Wolf20,Bray22,Wolf23} the N-terminal domain of Hsp90 bound to different inhibitors\cite{Wolf20,Bray22} or ions passing through the gramicidin ion channel\cite{Jaeger22}. In this study, on the other hand, we investigate the applicability of dcTMD for studying conformational changes of proteins. While we still want to bias the system along a simple one-dimensional coordinate $s$, the description of functional motion usually requires multiple CVs $\vec{x}$, in order to correctly resolve the reactive pathways as suggested by the free energy landscape $\Delta G(\vec{x})$.

To this end, we extend the theory of Ref.~\citenum{Post19}, which performs a dimensionality reduction to obtain appropriate CVs $\{x_j\}= \vec{x}$ and estimates the free energy $\Delta G(\vec{x})$ from dcTMD simulations, and derive a new estimate of the free energy based on the cumulant approximation.
Moreover, we discuss the calculation of the internal friction $\matrix{\itGamma}(\vec{x})$ of the protein,\cite{Best10,Soranno12,Schulz12,Echeverria14} which depends on the specific pathway on the free energy landscape as well as on the pulling velocity $v$ due to constraint-induced effects. \cite{Post22} Following Ref.~\citenum{Wolf23}, we finally calculate the pathway-specific transition rates associated with a given $\itGamma(s)$.

\begin{figure}[h!]
 \begin{centering}
   \includegraphics[width=0.48\textwidth]{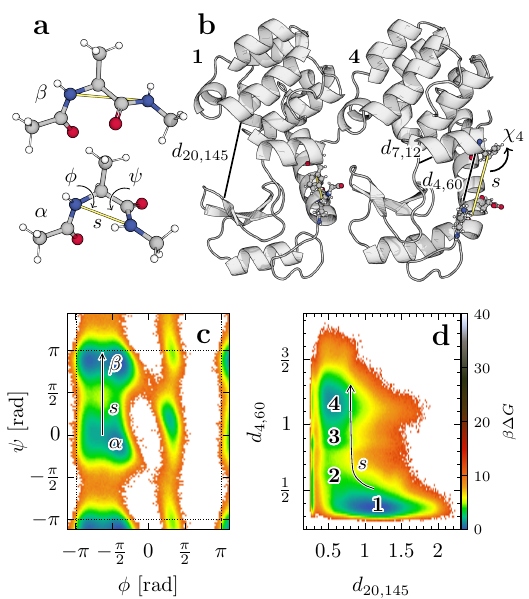}
\caption{Systems studied in this work. (a) Alanine dipeptide, showing the backbone dihedral angles $\phi$ and $\psi$, as well as the used pulling coordinate $s$. (b) Structure of the open state {\bf 1} and the closed state {\bf 4} of T4 lysozyme, showing the two main features $d_{20,145}$ and $d_{4,60}$ describing its open-closing transition, as well as $d_{7,12}$ and $\chi_{4}$ specified in the main text. (c) The free energy landscape of AlaD along $\phi$ and $\psi$ showing its main structural states. The arrow indicates the main pulling path, i.e., denotes the progress within reaction coordinate $s$. (d) The free energy landscape of T4L along $d_{20,145}$ and $d_{4,60}$.}
  \label{fig:protein_structures}
 \end{centering}
\end{figure}

To validate the estimation of these equilibrium fields from biased dcTMD simulations, we first consider the well-known alanine dipeptide (Ac-Ala-NHCH$_3$, for short AlaD) as a minimal model, showing transitions between its compact $\alpha$-state to its extended-$\beta$ state (Fig.~\ref{fig:protein_structures}a). The free energy landscape of AlaD is well described by a single pair of backbone dihedral angles $\phi,\psi$ and is shown in Fig.~\ref{fig:protein_structures}c. Putting our approach then to the test with the conformational change of a real protein, we consider the enzyme T4 lysozyme (T4L),\cite{Remington81,Dixon92,Yirdaw12}
which performs a hinge-bending motion between an open state and a closed state that can be described by two contact distances \cite{Ernst17,Post22a}, see Fig.~\ref{fig:protein_structures}b,d.

\section{Theory} 
\label{sec:theory}

%
%
\subsection{Free energy landscape}

In the case of pulling simulations along coordinate $s$, it was previously shown\cite{Hummer01a,Post19} that the free energy $\Delta G$ as a function of some CVs $\vec{x}$ can be calculated from Jarzynski's equation~\eqref{eq:free_energy_reaction_coordinate_exp} via
\begin{equation} \label{eq:free_energy_landscape_exponential}
\beta \Delta G(\vec{x}) = - \ln \left[ \upint \!\! \d s \cE{ \delta\!\left(\vec{x}-\vec{x}(\vec{q})\right) \, \euler^{-\beta W} }{s} \right] \; ,
\end{equation}
where $\vec{q} =(q_1,\dots,q_{3N})$ denote the atomic Cartesian MD coordinates of the system sampled along the pulling path $s$, i.e.  $\vec{q} = \vec{q}(s)$, and $\vec{x}(\vec{q})$ represent the resulting CVs.

Since the Jarzynski equation is known for notorious convergence issues of the biased exponential average, we again want to perform a cumulant expansion to second order as done in Eq.~\eqref{eq:free_energy_reaction_coordinate}. As detailed in SI Methods, this can be done by expanding Eq.~\eqref{eq:free_energy_landscape_exponential} in powers of $\beta$, giving
\begin{equation} \label{eq:free_energy_landscape_cumulant}
 \Delta G(\vec{x}) \approx \Delta \mathcal{G}(\vec{x}) + \cE{W}{\vec{x}} - \frac{\beta}{2} \cE{\delta W^2}{\vec{x}} \; .
\end{equation}
Here $\cE{\cdot}{\vec{x}}$ denotes an average over $\vec{x}$, for example,
\begin{align}
 \cE{W}{\vec{x}} &\coloneqq \frac{ \upint \!\! \d s \cE{ \delta\!\left(\vec{x}-\vec{x}(\vec{q})\right)W}{s}  }{ \upint \!\! \d s^\prime \cE{ \delta\!\left(\vec{x}-\vec{x}(\vec{q}) \right)}{s^\prime} } \; 
\end{align}
with $\delta W(\vec{x})= W(\vec{x})-\cE{W}{\vec{x}}$. Moreover we
introduced the nonequilibrium energy landscape \cite{Post19}
\begin{align}
 \beta \Delta \mathcal{G}(\vec{x}) &\coloneqq - \ln \left[ \upint \!\! \d s \cE{
\delta\!\left(\vec{x}-\vec{x}(\vec{q})\right) }{s} \right] \; , \label{eq:free_energy_landscape_biased}
\end{align}
which provides a meaning to the biased distribution of CVs.  Provided that the work distributions $\cE{W}{\vec{x}}$ resemble a Gaussian for every $\vec{x}$, we expect that the cumulant approximation \eqref{eq:free_energy_landscape_cumulant} provides a free
energy estimate with significantly improved convergence behavior than Eq.~\eqref{eq:free_energy_landscape_exponential}.

%
%
\subsection{Effective mass}

The Langevin equation \eqref{eq:langevin_equation} contains the mass
tensor $\matrix{M}$, which in general may depend on the CVs
$\vec{x}$. \cite{Schaudinnus16,Lee19} In the one-dimensional case, the CVs coincide with the
pulling coordinate, $\vec{x} \rightarrow s$, hence $\matrix{M} \rightarrow m$ 
is simply given by the reduced mass of the two (groups of) atoms
that are pulled apart along the distance $s$, and therefore need not to be
inferred from the data.
Alternatively, the effective mass $m$ can be
obtained from the equipartition theorem, e.g., $m \dot{x}^2 = 1/\beta$ per degree of freedom.
\cite{Vroylandt22} Assuming a diagonal mass matrix with elements
$M_{ii}$ depending on $\vec{x}$, we may calculate the expectation value
of the corresponding velocities $\dot{x_i}$ via an reweighted average,
\cite{Crooks00,Hummer05a} yielding
\begin{equation} \label{eq:mass}
	M_{ii}^{-1}(\vec{x}) =  \beta \, \frac{ \cE{\dot{x}_i^2 \,
            \euler^{-\beta W}}{\vec{x}} }{ \cE{\euler^{-\beta
              W}}{\vec{x}} } \approx \beta \,\cE{\dot{x}_i^2
        }{\vec{x}} \, . 
\end{equation}
Here the latter approximation holds if the bias of pulling does not affect
the velocity distribution, i.e., if we are still close to thermal
equilibrium.

%
%
\subsection{Friction}

To calculate the friction tensor $\matrix{\itGamma}$ as a function of
$\vec{x}$, we may generalize either Eq.\
\eqref{eq:friction_reaction_coordinate_exp} or
\eqref{eq:friction_reaction_coordinate_cum} to the multidimensional
case. Since the latter involves the tedious calculation of
autocorrelation functions of highly fluctuating
observables, we choose the former approach based on the dissipative
work $W_\text{diss} = W-\Delta G$. Exploiting the constraint
$s(t) = s_0 + v t$, we obtain for the friction along $s$
\begin{align}\label{eq:frics}
  \itGamma(s) = \frac{1}{v^2} \cE{\dot{W}_\text{diss}}{s},
\end{align}
which can be readily calculated from $\dot{W}_\text{diss}(t) = v f(t) - v\, \partial G / \partial s |_{s_0+vt}$.

To generalize to the calculation of $\matrix{\itGamma}(\vec{x})$,
it seems obvious to simply replace 
$\cE{\dot{W}_\text{diss}}{s}$ by $\cE{\dot{W}_\text{diss}}{\vec{x}}$.
Moreover, we need to account for the effect that the velocity
$\dot{\vec{x}}$ is not constant, while $\dot{s}=v$ is. To this end, we
focus on the important case, that the motions along $s$ and $\vec{x}$
are highly correlated, such that their average relaxation times given
by the quotients of the respective friction and mass are similar.
This consideration suggest the ansatz
\begin{align} \label{eq:friction_collective_variables2}
	\matrix{\itGamma}(\vec{x}) &= \frac{\matrix{M}(\vec{x})}{m}
  \cE{ \itGamma (s) }{\vec{x}}\nonumber \\
  &= \frac{\matrix{M}(\vec{x})}{m}
  \frac{1}{v^2} \cE{ \dot{W}_\text{diss} }{ \vec{x} },
\end{align}
where $m$ is the effective mass of the two atom groups
that are pulled apart along $s$, while $\matrix{M}(\vec{x})$
denotes the mass matrix as a function of $\vec{x}$.
Note that this expression resembles the diffusion tensor of overdamped Langevin dynamics reported in Refs.\ \citenum{Maragliano06,Evans22}, which also includes a time-scaling factor via $\itGamma(s)$.
Assuming that the friction matrix is diagonal with elements
$\itGamma_{ii}$, and employing the corresponding effective mass
$M_{ii}^{-1}(\vec{x})$ given by Eq.\ \eqref{eq:mass}, we obtain 
\begin{equation}
	\label{eq:friction_collective_variables}
	\itGamma_{ii}(\vec{x}) = \frac{\kb T}{m v^2}
        \frac{ \cE{ \dot{W}_\text{diss} }{ \vec{x} }}
        { \cE{\dot{x}_i^2 }{\vec{x}} } \; .
\end{equation}
Compared to Eq.~\eqref{eq:frics}, the friction is given by the
replacements $\cE{\dot{W}_\text{diss}}{s} \rightarrow
\cE{\dot{W}_\text{diss}}{\vec{x}}$ and
$v \rightarrow \cE{\dot{x}_i^2 }{\vec{x}}$ with the additional
factor of $(\kb T) / (m v^2)$, which corrects for the biased pulling
velocity violating the equipartition theorem.
In SI Methods, we also discuss alternative ways to obtain
$\matrix{\itGamma}(\vec{x})$.

\section{Systems and Methods} 
\label{sec:methods}

%
%
\subsection{Alanine dipeptide}

\subparagraph{Molecular dynamics simulations}

Unbiased simulations of alanine dipeptide were previously reported on and analyzed in Ref.~\citenum{Nagel19}. In brief, the peptide's dynamics were simulated using Gromacs v2016.3 \cite{Abraham15} employing the Amber99SB-ILDN force field\cite{LindorffLarsen10} in a solvated dodecahedral box with image distance of 2.71~nm with 452 TIP3P water molecules at $T=300$~K, using the Bussi thermostat\cite{Bussi07} ($\tau=0.1$~ps). Electrostatics were handled using PME\cite{Darden93}. Both the minimal Coulomb real-space and the van der Waals cutoffs were set to 1.2~nm. Covalent bonds involving hydrogen atoms were constrained with SHAKE\cite{Ryckaert77}. Production runs comprised 500~ns of simulation in 20 fs resolution.\cite{Nagel19}

\subparagraph{Targeted MD simulations}

Aiming to enforce the $\alpha \rightarrow \beta$ transition by means of constraining a suitable coordinate, we employ the Gromacs PULL code. Limited to only distances, we use the distance $s$ between the two nitrogen atoms within the peptide as reaction coordinate to mimic the switch of the dihedral angle $\psi$ (Fig.~\ref{fig:protein_structures}a). In this way, we are studying how a sub-optimal reaction coordinate influences the quality of estimates from the biased transition. For more complex systems, this is usually the case, when the optimal reaction coordinate is hard to define.

Starting from $s_0=0.26$~nm, we prepared 1000 starting configurations each, sampled from the equilibrium data, which were in turn equilibrated again with fixed $s$ for 1~ns. Then, using a range of different pulling velocities $v$, each of the 1000 resulting structures were subjected to constant velocity pulling with $s (t) =s_0+vt$ until $s=0.39$~nm. We save the applied constraint forces $f$ every femtosecond and compute the work $W$ [Eq.~\eqref{eq:work}] by integration via trapezoidal rule, where it is only necessary to store up to 10000 frames $W(s)$ per trajectory.
%
%

\subparagraph{Estimation of masses}

To calculate the friction tensor $\matrix{\itGamma}(\phi, \psi)$ from
Eq.~\eqref{eq:friction_collective_variables2}, we need to estimate the
mass $m$ representing the effective mass of the two atom groups that
are pulled apart along $s$, as well as the mass tensor
$\matrix{M}(\phi, \psi)$ accounting for the effective moment of
inertia associated with dihedral angles $(\phi, \psi)$, respectively.
Naively, we may take for $m$ the reduced mass of the
pulled atom group, i.e., $m=7$~g/mol for the two nitrogen atoms.
Since these atoms are strongly coupled to the rest of the solvated molecule via covalent bonds,
however, the true effective mass can be significantly larger. 
%
The diagonal elements of the mass tensor $\matrix{M}(\phi, \psi)$ were
estimated using Eq.~\eqref{eq:mass}. As shown in
Fig.~\mbox{\SIFigAlaMass}, we obtained the average values
$\matrix{M}_\phi = 0.031$ and $\matrix{M}_\psi = 0.038$~g
nm$^2$/mol/rad$^2$.

\subparagraph{Langevin simulations}

To obtain an equilibrium estimate of the friction, we solved Langevin
equations (\ref{eq:langevin_equation}) for $\vec{x} = s$ and for
$\vec{x} = (\phi,\psi)$, using the corresponding free energy
landscapes from the unbiased MD data and the integrator of Bussi and
Parrinello\cite{Bussi07}. 
As the Langevin dynamics of AlaD is clearly overdamped, it does not
depend on the choice of the mass. 
Here we used the reduced mass of the two pulled nitrogen atoms (i.e., $m=7$~g/mol).
By searching for a constant friction factor
that recovers the $\alpha \leftrightarrow \beta$ transition time of
the the unbiased simulations, we obtained
$\itGamma = 12\, 750$~g/mol/ps for $\vec{x} = s$ and $\itGamma =13$~g
nm$^2$/mol/ps/rad$^2$ for $\vec{x} = (\phi,\psi)$.

%
%
\subsection{T4 lysozyme}

\subparagraph{Molecular dynamics simulations}

Unbiased simulations were previously performed by Ernst et al.\cite{Ernst17} In brief, the system consists of the 154-residue protein, 8918 TIP3P water molecules, 27 Na$^+$ and 35 Cl$^-$  ions in a triclinic box, using the Amber ff99SB$^*$-ILDN force field\cite{LindorffLarsen10} with SHAKE constraints applied to all bonds involving hydrogen atoms. Electrostatics were handled using PME\cite{Darden93}. Both minimal real-space Coulomb and van der Waals cutoffs were set to 1.2~nm. All runs used a time step of 2~fs. After an initial steepest descent minimization, the system was equilibrated for 10~ps in $NVT$. After that, a $1$~ns $NPT$ equilibration followed, then $5$~ns without position restraints. Then, the box was re-scaled to the average box volume, resulting in a $6.96 \times 6.68 \times 6.34$~nm$^3$ box. This structure was again equilibrated for 10~ns in $NVT$. This equilibrated structure together with atomic velocities formed the input for a 61~\us\ long $NVT$ simulation, saving atomic positions every picosecond.

\subparagraph{Targeted MD simulations}

As biasing coordinate $s$, we here chose the distance between the center of mass of all carbon atoms of residue 4 and the center of mass of residue 60 and 64 (Fig.~\ref{fig:protein_structures}b), which we determined earlier to trigger the open-closed conformational change in T4L\cite{Ernst17}. To generate seeds for simulations pulling from open state $\textbf{1}$ to the closed state $\textbf{4}$, 50 frames at $s=0.38$~nm were taken from the unbiased trajectory. With fixed $s$, the structures were equilibrated for $100$~ns. Taking a structural snapshot each 5~ns generated 20 input structures per simulation. These 1000 statistically independent structures were used as input for constant velocity pulling simulations. 

To ensure that the pulling motion during the enforced open--closed transition remained the slowest process and that all orthogonal degrees of freedom had sufficient time to relax, we used an adaptable pulling velocity scheme that was dependent on the range in $s$. We started pulling with $v=-0.05$~m/s from $s=0.3$ to $0.38$~nm, with $v=0.02$~m/s from $s=0.38$ to $0.5$~nm, with $v=0.01$~m/s from $s=0.5$ to $0.7$~nm, continuing with $v=0.02$~m/s from $s=0.7$ to $1.1$~nm and finally with $v=0.05$~m/s from $s=1.1$ to $1.7$~nm. Atomic positions were saved with 1000 frames per 0.1~nm.

\subparagraph{Langevin simulations}

Similar as described above for AlaD, we ran a Langevin simulation to obtain an 
equilibrium reference value for the friction $\itGamma(s)$. As we found that the T4L dynamics are highly overdamped (see below), we furthermore employed an integration of an overdamped Langevin equation via a first-order Runge-Kutta integrator.

\section{Results and Discussion} 
\label{sec:results}

%
%
\subsection{Alanine dipeptide}

\subparagraph{Estimating the free energy landscape}

\begin{figure}
	\begin{centering}	
		\includegraphics[width=0.48\textwidth]{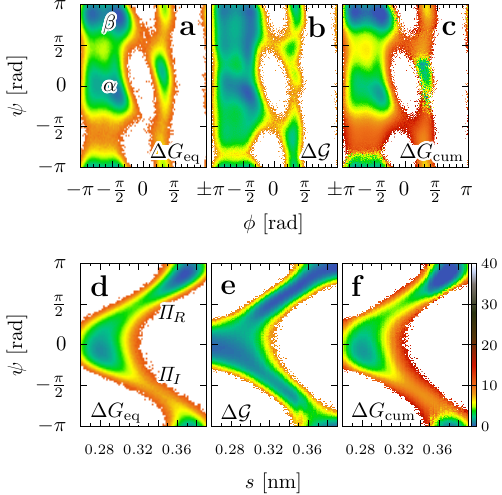}
		\caption{Energy landscape of AlaD, given in units of $\kb T$. (a-c) Landscape in dihedral angles $\phi$ and $\psi$. (d-f) Landscape in $s$ and $\psi$. Compared are the equilibrium free energy $\Delta G_\text{eq}$ (a,d), the biased energy landscape $\Delta \mathcal{G}$ (b,e) and the corresponding reweighed free energy $\Delta G_\text{cum}$ (c,f).} 
		\label{fig:free_energy_alaD}
	\end{centering}
\end{figure}

To benchmark our approach of estimating multi-dimensional free energies $\Delta G (\vec{x})$ and friction fields $\matrix{\itGamma}(\vec{x})$, we employ the well-understood AlaD system in water. The most prominent structural conformations of AlaD are its compact $\alpha$ state and its extended $\beta$ state.
They can be distinguished by the two backbone dihedral angles $\phi$ and $\psi$, and are often visualized by the so-called Ramachandran plot (Fig.~\ref{fig:free_energy_alaD}a), which shows the free energy estimate $\Delta G_\text{eq}$ from the unbiased MD simulations. 

The structural change $\alpha\rightarrow\beta$ via a change in the positive direction of $\psi$ (see Fig.~\ref{fig:protein_structures}c) describes the strongly preferred, or ``regular'', reaction pathway $\itPi_R$. The change along the negative direction via the ``irregular'' pathway $\itPi_I$ is disfavored due to steric hindrance by the CH$_3$ side-chain around $\psi \approx - 2 \pi / 3$. Along $\phi \approx -\pi / 2$, Fig.~\ref{fig:free_energy_alaD}a shows a split of $\alpha$ (as well as $\beta$) into two close basins with small barrier, which results in short transition times and thus are of no further interest here. Additionally, a lowly populated left-handed conformation exists at $\phi\approx 0.3\pi$,
which is not of interest for our investigation.

Having a general understanding of AlaD's unbiased conformational distributions, we consider the case of pulling the system along coordinate $s$ to trigger a transition from the $\alpha$ to the $\beta$ state. Using a fixed pulling velocity $v = 0.1$~m/s, we constrain the inter-nitrogen distance $s$ instead of the dihedral angle $\psi$ (Fig.~\ref{fig:protein_structures}a).
We chose this approach because distance constraint-based pulling is available in Gromacs via the Shake algorithm\cite{Ryckaert77}, while dihedral pulling is not implemented. \footnote{We note that dihedral constraint pulling is theoretically possible, see, e.g., Ref.~\citenum{Fixman74,Sprik98,Pechlaner22}.}
In this way, we can study how propagating $s$ triggers a change in $\psi$, which is particularly relevant in systems where a one-dimensional reaction coordinate is not able to describe the considered functional processes. 

First, we consider the biased energy landscape $\Delta\mathcal{G}$ using Eq.~\eqref{eq:free_energy_landscape_biased}, where from Fig.~\ref{fig:free_energy_alaD}b we can see a significant lowering of the energy in the transition region. After applying reweighting using Eq.~\eqref{eq:free_energy_landscape_cumulant}, all free energy minima are again clearly visible (Fig.~\ref{fig:free_energy_alaD}c). Fig.~\mbox{\SIFigAlaConvergence} displays a comprehensive convergence analysis of the free energy estimates. Note that our cumulant-based estimator introduces a small bias towards the $\beta$ state due to deviations of the work distribution from a Gaussian, which is also obtained by the exponential average (Fig.~\mbox{\SIFigAlaReverse}). We also simulated the back-transition $\beta \to \alpha$, yielding similar results, see Fig.~\mbox{\SIFigAlaReverse b}.

The correspondence between $s$ and $\psi$ can be visualized by their joint distribution, or the corresponding free energy landscape in Fig \ref{fig:free_energy_alaD}d, showing in more detail the split between the native path $\itPi_R$ and the irregular path $\itPi_I$. When we pull the peptide along $s$, we clearly see that the trajectories follow these two pathways from $\alpha$ towards $\beta$ (Fig \ref{fig:free_energy_alaD}e). Interestingly, with the bias, a considerable number of trajectories (20\%) choose the otherwise disfavored direction (see Fig.~\mbox{\SIFigAlaPathratio} for details). This bias is again correctly accounted for by the reweighting procedure (Fig \ref{fig:free_energy_alaD}f), showing that $s$ is indeed a useful substitute for $\psi$.

\subparagraph{Evaluating friction profiles}

We first consider the friction $\itGamma(s)$ obtained from the change in dissipated energy, calculated via Eq.~\eqref{eq:friction_reaction_coordinate_cum}. Since the partition into two distinct pathways results in a bimodal work distribution at fixed $s$ (see Fig.~\mbox{\SIFigAlaConvergence a,b}) we need to perform separate friction analyses for each path.\cite{Wolf23}. Here we here focus on the dominant path $\itPi_R$, and also exclude trajectories transitioning to a left-handed conformation.

\begin{figure}
	\begin{centering}	
		\includegraphics[width=0.45\textwidth]{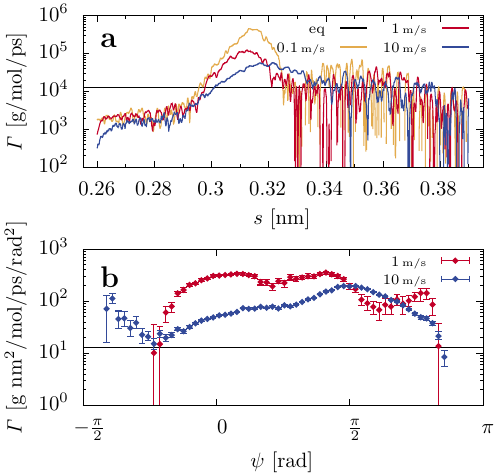}
		\caption{Friction of AlaD along the the regular pathway $\itPi_R$. (a) Profile in $s$ comparing three pulling velocities $v$. Data are averaged over 0.001~nm. (b) Friction along the backbone dihedral angle $\psi$, estimated for $v=10$~m/s. Error bars indicate 95\% confidence intervals estimated by bootstrapping. Black straight lines in both panels correspond to a constant friction resulting in transition times in equilibrium Langevin simulations that are comparable to the ones in unbiased MD simulations.}
		\label{fig:friction_alaD}
	\end{centering}
\end{figure}

 Figure \ref{fig:friction_alaD}a shows the resulting friction $\itGamma(s)$ along $\itPi_R$ and compares its positional dependency for three different constraint velocities: slow $v=0.1$~m/s (yellow), intermediate $v=1$~m/s (red), and fast $v=10$~m/s (blue). 
Apart from the constrained-induced peak of $\itGamma(s)$ around $s \approx 3.1$~nm (see below), all velocities result in comparable friction profiles. Note that at large $s$, the friction is subject to large fluctuations due to accumulation of noise and the log-scale representation.
Also shown is the friction, $\itGamma_{\rm eq} = 12\, 750$~g/mol/ps (black), which was obtained from a simplified Langevin equation using the unbiased free energy profile and assuming constant friction (see Methods). Compared to previous results obtained for ligand-protein unbinding, \cite{Wolf20,Bray22,Wolf23}
the overall magnitude found for the internal friction seems quite high, which indicates the existence of several strongly interacting degrees of freedom, which are orthogonal to the pulling coordinate $s$. 

By comparing various velocities, we may identify potential biases arising from specific reaction paths that are only accessible under constraint pulling. In case of slow velocities ($v=0.1$~m/s), from about $0.3$ to $0.32$~nm, there is a clear rise in friction of at least one order of magnitude. This region corresponds to the junction between $\itPi_R$ and $\itPi_I$ in the energy landscape shown in Fig.~\ref{fig:free_energy_alaD}e. At this slow pulling velocity and due to the resulting pseudo-stationarity imposed by the constraint, the system has sufficient time to switch between paths, which
results in an overestimation of the friction coefficient\cite{Wolf23}. In contrast, for faster velocities like $v=1$~m/s, the peak starts to vanish due a rapidly enforced commitment to one path, which seems to correspond better to the situation in the unbiased case. For the even faster $v=10$~m/s, the constraint causes an overbending of the N-C$_\upalpha$-C bond-angle that leads to a peak shift to larger $s$. 

Since the pulling coordinate $s$ is only an indirect way
to describe the dynamics of AlaD, it is interesting to consider
the friction as a function of the main reaction coordinate $\psi$.
Calculating $\itGamma(\psi)$ from
Eq.~\eqref{eq:friction_collective_variables}, Fig.\
\ref{fig:friction_alaD}b shows the resulting friction
obtained for pulling velocities $v=1$ and $10$~m/s. 
As for $\itGamma(s)$, the two cases are found to approximately coincide when the system is in the two metastable states $\alpha$ ($\psi \approx - \pi/4$) and $\beta$ ($\psi \approx 3\pi/4$), but deviate in the transition region where the friction is higher for $v=1$\,m/s.
The overall rise at $\psi \approx 0$ and drop at $\psi \approx \pi/2$ of  $\itGamma(\psi)$ can be explained via the inspection of the joint distribution of $s$ and $\psi$ (Fig.~\ref{fig:free_energy_alaD}d) and the friction profile $\itGamma(s)$ above.
Starting pulling at $\psi \approx -\pi/4$, the friction increases
when the system approaches the junction of the two paths $\itPi_R$ and $\itPi_I$.
At $\psi \approx \pi/2$, all trajectories have committed to path $\itPi_R$, which results in a drop of the friction.
The dcTMD results are found to overestimate the results from a equilibrium Langevin model assuming constant friction (see Methods), in particular in the transition region. As noted above, this discrepancy may be related to the fact that we assumed a lower limit for the effective mass $m$ of the pulling, which directly enters the calculation of $\itGamma(\psi)$ in Eq.~\eqref{eq:friction_collective_variables}.

Employing the dcTMD results for the free energy $\Delta G (s)$ and the
friction $\itGamma(s)$ to run Langevin simulations (see Methods), we
may finally estimate the kinetic rates of the $\alpha\leftrightarrow\beta$
transitions. In the $\alpha\rightarrow\beta$ pulling direction, the
dcTMD rate $k_{\alpha\rightarrow\beta} =1$/(42\,ps) is 37\,\% higher
than the MD reference result of 1/(67\,ps), which represents a quite
good accuracy for the estimation of a kinetic rate. 
In the backwards direction $\beta\rightarrow\alpha$, we obtain 
$k_{\beta\rightarrow\alpha} =1$/(272\,ps) from dcTMD, which
underestimates the MD result of 1/(118\,ps) by roughly a factor 2.


%
%
\subsection{T4 lysozyme}

To put the applicability of our method to the test with a conformational transition of a full protein, we investigate the open-closed transition of T4L (see Fig.~\ref{fig:protein_structures}b). In previous works, we have gained a detailed understanding of the residue-wise dynamics that define this transition\cite{Ernst17,Post22a}: the respective reaction pathway is defined by a cooperative contact network that connects the protein's active side (aka the ''mouth'' area) around the distance pair $d_{20,145}$ and the opposite region of the protein around Phe4 (the ''jaw joint'' or ''hinge'' area) described by the distance $d_{4,60}$. 
Fig.~\ref{fig:protein_structures}d shows the free energy landscape along these coordinates, which exhibits 4 states that we describe in the following.

The open-closed transition \textbf{1}$\to$\textbf{4} starts in the open state \textbf{1} with a closing attempt toward state \textbf{2}, where $d_{20,145}$ decreases, but Phe4 is still buried in its pocket. This structural change is transported via rigid contacts within the N- and C-terminal domains to an ensemble of hydrophobic residues, i.e., the ''hydrophobic core'', which causes the transition into the pre-closed state \textbf{3} by a change of the Phe4 side chain from a hydrophobically-buried to a solvent-exposed state (monitored by the contact distance $d_{4,60}$). The final closed state \textbf{4} is further stabilized by a piston-like motion of helix $\alpha_1$ (characterized by an opened $d_{7,12}$) and a final rotation of the Phe4 side chain around its side chain dihedral angle $\chi_4$ into the bulk water, which appears as an additional increase of $d_{4,60}$. A detailed description of this long-range allosteric coupling is given in Ref.~\citenum{Post22a}.

While the original unbiased trajectory of 61~\us\ simulation time provides an exceptional level of dynamical details, it still only includes about ten full-circle \textbf{1}$\leftrightarrow$\textbf{4} transitions, which nevertheless indicate a good agreement with the experimentally observed waiting time\cite{Sanabria20} of $\approx 4$~\us. Moreover, Ernst et al.\cite{Ernst17} demonstrated that constraint-based pulling of the Phe4 side chain triggers the \textbf{1}$\to$\textbf{4} transition. This previous study utilized the distance between the center of mass of the Phe4 side chain and the combined center of mass of residues Lys60 and Glu64 (Fig.~\ref{fig:protein_structures}b) as pulling coordinate $s$. Note that pulling the Phe4 side chain out of its hydrophobically buried position into the solvent is the only practical option to enforce the open-close conformational transition of T4L\cite{Post22c}: the reverse reaction, i.e., pulling the side chain into the protein, requires to establish the correct packing near-order, which is beyond the time scales desired in our biased simulations and is well understood in the case of protein--ligand complexes\cite{Wolf23a}. We use this conformational transition as a prime example for the construction of a Langevin model via dcTMD. 

\subparagraph{Free energy landscape}

\begin{figure}[ht!]
	\begin{centering}	
		\includegraphics[width=0.48\textwidth]{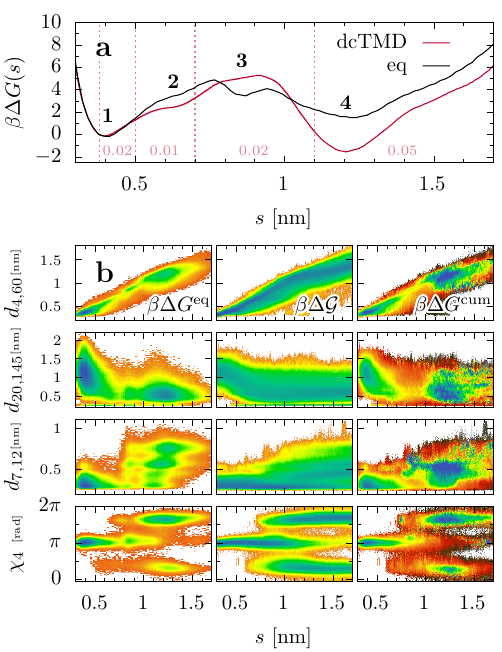}
		\caption{Free energy of T4 lysozyme. (a) dcTMD Free energy profile along pulling coordinate $s$, compared to the unbiased MD simulation. Vertical lines indicate regions of different velocity $v$, as indicated on the bottom in units of m/s. (b) Two-dimensional free energy as a function of $s$ and various relevant internal coordinates, comparing unbiased, biased and reweighted cases (see Fig.~\ref{fig:protein_structures}b for the positions of indicated distances within the protein).}
		\label{fig:free_energy_t4l}
	\end{centering}
\end{figure}

We begin our discussion by characterizing the conformational states in terms of $s$ only, based on the corresponding free energy landscape $\Delta G(s)$, see Fig.~\ref{fig:free_energy_t4l}a. Here, we compare the unbiased distribution of states (black) to the distributions predicted by the biased pulling simulations. Focusing first on the unbiased results, we identify the above described states \textbf{1} to \textbf{4} by distinct regions separated by free energy barriers. For small $s$ we find the open state \textbf{1}, where at large $s$ we have the closed state \textbf{4}, with the less pronounced short-lived intermediate state \textbf{3} in between, while \textbf{2} only appears as a shoulder in the projection onto $s$.
The free energies associated with theses states are compared to the results obtained from the pulling simulations (red), where the constraint velocity $v$ ranges between $0.05$ and $0.01$~m/s, adjusted to an appropriate value for regions in $s$ according to conformational states, to allow for sufficient relaxation of all orthogonal degrees of freedom (see Methods). 
Note that under this variable velocity scheme, the work distribution clearly appears to follow a Gaussian shape (Fig.~\mbox{\SIFigLysConvergence a,b}), justifying the cumulant approximation Eq.~\eqref{eq:free_energy_reaction_coordinate_cum}. 

Starting with the \textbf{1} $\rightarrow$ \textbf{2} transition, the pulling simulations yield a lower free energy estimate of state \textbf{2} than the equilibrium MD by $\approx 1~\kb T$. The exposure of the Phe4 ring occurring during transition \textbf{2} $\rightarrow$ \textbf{3}, on the other hand, is estimated to have a higher energy barrier, where \textbf{3} disappears as separate state. Lastly, while the \textbf{3} $\rightarrow$ \textbf{4} barrier is larger than in the unbiased case, the free energy of state \textbf{4}  is estimated significantly lower, differing by about 3~$\kb T$ from the unbiased value. 
As a consequence, the dcTMD calculations predict an equilibrium occupation of 34 and 66\% for the open and the closed state, while the unbiased simulations gave 70 and 30\%, respectively.
Despite these difference in the free energy profile to the unbiased calculation, we think that the biased calculation might actually represent a better estimate of the free energy of the closed state \textbf{4}, because the sampling of the dcTMD simulations is much better than the rather limited sampling in the unbiased simulations. Indeed, Fig.~\mbox{\SIFigLysConvergence c} indicates a reasonably converged free energy estimate from pulling simulations. We note that correctly reproducing the T4L free energy profile poses a challenge for computational predictions of free energies, as the typical mean error of free energy calculation methods\cite{Fu22,Wolf23a} is on the order of $\sim 3 \kb T$.

Keeping the sampling issues in mind, we qualitatively analyze the detailed transition mechanism based on our estimates $\Delta G^{\rm cum} (\vec{x})$ with $\vec{x} = (d_{4,60},\, d_{20,145},\, d_{7,12},\, \chi_4)$ and compare them to the respective unbiased free energy profiles $\Delta G^\text{eq}$ from Ref.~\citenum{Post22a}.
Fig.~\ref{fig:free_energy_t4l}b shows how the pulling in $s$ results in a change of these important features. In general, the nonequilibrium energy landscapes $\Delta \mathcal{G}$ indicates a uniform sampling along the four investigated collective variables, which connects all relevant regions with minima in the free energy landscape. An exception is a local minimum in $d_{7,12}$ at $\approx 0.75$~nm that is not reached. Applying a bias along $s$ therefore seems to trigger most of the conformational changes of relevance. That is, by calculating the reweighted free energy via the dissipation-correction, we qualitatively recover the positions of local free energy minima: for the distance $d_{4,60}$ parameterizing the movement of the Phe4 side chain, we find the hydrophobically buried and solvent-exposed conformations at small and large $s$, respectively. In the case of $d_{20,145}$, the positions of the ''mouth-open'' (small $s$) and ''mouth-closed'' (large $s$) conformations clearly emerge. The agreement is worse for $d_{7,12}$, which recovers a minimum in free energy for small $s$, but contracts several sub-minima in $d_{7,12}$ at long $s$ into a single minimum at too small $d_{7,12}$. Fig.~\mbox{\SIFigLysConvergence d} proposes that this issue comes from bad coverage of $d_{7,12}$ resulting in an underestimated $\E{\delta W^2}$ and thus $W_{\rm diss}$. Lastly, the significant minima in free energy for $\chi_4$ are recovered by the dissipation correction, with exception of a shallow minimum at $(s,\chi_4) \approx (0.85 \text{ nm}, \pi)$ that could be mistaken as noise after correction. In summary, the reweighted landscape $\Delta G^\text{cum}$, though not fully converged, is capable to recover most of the conformational details that appear in a \textbf{1} $\rightarrow$ \textbf{4} transition without bias, albeit without the quantitative reproduction of the depths of the free energy minima.

\subparagraph{Friction analysis}

\begin{figure}
	\begin{centering}
		\includegraphics[width=0.38\textwidth]{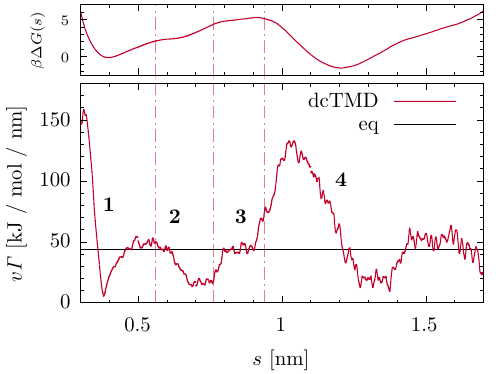}
		\caption{Friction profile of T4L estimated from dcTMD simulations with the dcTMD free energy profile above for reference. Shown are the product of friction $\itGamma$ and pulling velocity $v$, i.e., $\frac{\d W_\text{diss}}{\d s}$. Dashed vertical lines indicate the transition regions between the meta-stables states \textbf{1} to \textbf{4}. Data are smoothed by a window average over $0.01$~nm. The black line indicates a reference value (multiplied with the minimal velocity) for Langevin equation simulations that reproduces \textbf{1} $\leftrightarrow$ \textbf{4} waiting times from unbiased MD simulations.}
		\label{fig:friction_t4l}
	\end{centering}
\end{figure}

Completing the picture by estimating the friction profile along $s$, we start the discussion with the non-equilibrium friction estimate of T4L using Eq.~\eqref{eq:friction_reaction_coordinate_cum}. Because of our adaptive scheme for the constraint velocity, we consider in Fig.~\ref{fig:friction_t4l} the product $v\itGamma$ (i.e., the friction force $\frac{\d W_\text{diss}}{\d s}$), in order to preserve the continuity of the profile. 
%
Starting to pull from a position of minimal free energy at $s=0.38$~nm (see Fig.~\ref{fig:free_energy_t4l}), we first probe the behavior of pulling to shorter distances for the sake of a complete picture over the full investigated range of $s$. The resulting peak in friction (that coincides with a sharp increase of the free energy) reflects the excitation of degrees of freedom that resist a motion of Phe4 deeper into the hydrophobic core, which is of no further relevance here. 

Pulling from state\,\textbf{1} at $s = 0.38$~nm to larger values, the friction profile varies between $\sim 10$ and 130 kJ/mol/nm and exhibits several maxima, which arise from the interaction with degrees of freedom that are orthogonal to $s$. In particular, the first maximum at $s\sim 0.5\,$nm most likely results from a salt bridge between Glu5 and Lys60 that buries Phe4 within the protein core,\cite{Post22a} and which needs to open for the \textbf{1}$\to$\textbf{2} transition. After this opening, $v \itGamma$ decreases to a minimum that persists up to $s \approx 0.76$~nm. The following increase of $v \itGamma$ reflects Phe4 needing to pass across Phe67 in the \textbf{2}$\to$\textbf{3} transition. As displayed in Fig.~\ref{fig:free_energy_t4l}b, this change corresponds to the occurrence of two additional free energy minima along $\chi_4$.
The subsequent major maximum of $v\itGamma$ from $s \approx 0.9$~nm  up to $s \approx 1.2$~nm is due to the final rotation of $\chi_4$ during the \textbf{3}$\to$\textbf{4} transition, which corresponds to the vanishing of the free energy minimum at $\chi_4 \sim \pi$ and the final transition into $\chi_4 \sim \pi/2, 3\pi/2$, see Fig.~\ref{fig:free_energy_t4l}b. For $s \gtrsim 1.4$~nm, friction increases again due to over-stretching of the Phe4 side chain. 

Overall the friction profile compares well to the result $v\itGamma(s) = 4.4\times 10^6$~g/(mol ps) (black line) from an equilibrium Langevin equation, which
 is able to reproduce the open-closed relaxation time $\tau_{\textbf{1} \leftrightarrow \textbf{4}} \approx 1.8$~\us\ of the unbiased MD, see Fig.~\mbox{\SIFigLysFriction c,d}. We note that this value for the internal friction is quite high, because it reflects the multitude of orthogonal degrees of freedom within the protein involved in the 
 cooperative change of an extended contact network during the open-close transition of T4L\cite{Post22a} that strongly couple with our chosen coordinate $s$.

 Calculating the friction for various pulling velocities, Fig.\ \SIFigLysFriction{b} shows that we obtain a similar shape of the friction profiles $\itGamma(s)$, as well as similar average values, i.e., $v \itGamma =$51, 68 and 64 kJ/mol/nm for $v=$
 0.01, 0.025 and 0.05 m/s, respectively. That is, we find a Coulomb-type behavior of the velocity dependence of the friction, $\itGamma \propto 1/v$, which is known, e.g., from slip-stick friction exhibited by the Prandtl-Tomlinson model.\cite{Mueser11} 

Using the dissipation-corrected free energy $\Delta G^{\rm cum} (s)$ and the non-equilibrium friction profile $\itGamma(s)$ to perform long Langevin simulations, 
we obtain an overall relaxation time $\tau_{\textbf{1} \leftrightarrow \textbf{4}} \approx 3$~\us, which is in excellent agreement with the experimentally determined $\approx 4$~\us. More specific, we obtain the rates $k_{1\rightarrow 4} \approx 0.3$~\us$^{-1}$ (compared to $0.2$~\us$^{-1}$ in the unbiased MD) and $k_{4\rightarrow 1}^\text{TMD} \approx 0.1$~\us$^{-1}$ ($ 0.5$~\us$^{-1}$ in the unbiased MD), which also reflects the population ratio 34/66 of state {\rm 1} and {\rm 4} (70/30 in the unbiased MD). 


\section{Concluding remarks} 
\label{sec:conclusion}
We have outlined a strategy to use dcTMD pulling simulations to study
rare conformational transitions in proteins. Since the description of
functional motion typically requires a reaction coordinate $\vec{x}$
that is not identical with the pulling direction, we have derived
explicit expressions to calculate the free energy landscape
$\Delta G(\vec{x})$ as well as the friction
$\matrix{\itGamma}(\vec{x})$ from dcTMD. 
The approach was
tested by studying the $\alpha$-$\beta$ transition of AlaD as a
proof-of-principle model, and the open-closed transition of T4L as a
challenging example.

We have shown that dcTMD greatly facilitates the sampling of rare
transitions, which leads to an improved estimate of the equilibrium
free energies of the two main metastable states of T4L as well as of
the transition rates between them, both being in excellent
agreement with experiment. Moreover we have demonstrated that the
friction profile $\matrix{\itGamma}(\vec{x})$ yields valuable information on
the dynamics of degrees of freedom that are orthogonal to the reaction
coordinate, which provides an alternative approach to interpret the 
mechanism underlying the biomolecular process.

%
%
\begin{acknowledgement}

We thank Kerstin Falk (Fraunhofer IWM Freiburg), Tanja Schilling and Fabian Koch (University of Freiburg), and Benjamin Lickert (Fraunhofer EMI Freiburg) for helpful discussions.
This work has been supported by the Deutsche Forschungsgemeinschaft (DFG) via the Research Unit FOR 5099 ``Reducing complexity of nonequilibrium''(project no. 431945604).
The authors acknowledge support by the bwUniCluster computing initiative, the High Performance and Cloud Computing Group at the Zentrum f\"ur Datenverarbeitung of the University of T\"ubingen, and the Rechenzentrum of the University of Freiburg, the state of Baden-W\"urttemberg through bwHPCand the DFG through grants no. INST 37/935-1 FUGG and nohttps://de.overleaf.com/project/64c10bab4aa4d319adc332bc. INST 39/963-1 FUGG.

\end{acknowledgement}

%
%
\subsection*{Data Availability Statement}

Unbiased simulations of alanine dipeptide were previously reported on and analyzed in Ref.~\citenum{Nagel19}, and unbiased T4 lysozyme simulations were previously performed by Ernst et al.\cite{Ernst17} Biased simulation trajectories are available from the authors upon reasonable request.

%
%
\begin{suppinfo}

Derivation of Eq.~\eqref{eq:free_energy_landscape_cumulant} and an alternative way to obtain the friction coefficient $\matrix{\itGamma}(\vec{x})$ via a fit to the dissipated work. Figures of mass and friction estimations, work distributions, and the convergence behavior obtained from dcTMD simulations of AlaD and T4L. (PDF)

\end{suppinfo}


\providecommand{\latin}[1]{#1}
\makeatletter
\providecommand{\doi}
  {\begingroup\let\do\@makeother\dospecials
  \catcode`\{=1 \catcode`\}=2 \doi@aux}
\providecommand{\doi@aux}[1]{\endgroup\texttt{#1}}
\makeatother
\providecommand*\mcitethebibliography{\thebibliography}
\csname @ifundefined\endcsname{endmcitethebibliography}
  {\let\endmcitethebibliography\endthebibliography}{}


\newpage
\onecolumn

\setcounter{figure}{0}

\section*{Supplementary Information}

\section*{Supplementary Methods}

\subparagraph*{Cumulant approximation of the free energy}

In analogy of the cumulant approximation of Jarzynski's equation
\begin{align}
 \beta \Delta G(s) &=  - \ln \cE{ \euler^{-\beta W} }{s} \\
 &\approx  \cE{ W }{s} - \frac{\beta}{2}  \cE{ \delta W^2 }{s} \, ,
\end{align}
we here derive an analogous expansion of the free energy landscape (Eq.~6 of the main text)
\begin{equation}
	\beta \Delta G(\vec{x}) = - \ln \left[ \upint \!\! \d s \cE{ \delta_\vec{x} \, \euler^{-\beta W} }{s} \right] \; ,
\end{equation}
with shorthand $\delta_\vec{x} = \delta(\vec{x}-\vec{x}(q))$. We first expand the exponential function $\euler^{-x}=1-x+\tfrac{x^2}{2}-\tfrac{x^3}{6}+\dots$ in powers of $\beta$,
\begin{align}
	\beta \Delta G(\vec{x}) &= - \ln \left[ \upint \!\! \d s \cE{ \delta_\vec{x} \, \left( 1 -\beta W + \tfrac{\beta^2}{2} W^2 - \dots \right) }{s}  \right] \\
	&= - \ln \left[ \upint \!\! \d s  \cE{ \delta_\vec{x} }{s} - \beta \upint \!\! \d s \cE{ \delta_\vec{x} W}{s} + \tfrac{\beta^2}{2} \upint \!\! \d s \cE{ \delta_\vec{x} W^2 }{s} - \dots   \right] \\
	&= - \ln \left[ \left( \upint \!\! \d s  \cE{ \delta_\vec{x} }{s} \right) \times \left( 1 - \beta \frac{ \upint \!\! \d s \cE{ \delta_\vec{x} W}{s}}{ \upint \!\! \d s  \cE{ \delta_\vec{x} }{s}} + \frac{\beta^2}{2} \frac{ \upint \!\! \d s \cE{ \delta_\vec{x} W^2 }{s}}{ \upint \!\! \d s  \cE{ \delta_\vec{x} }{s}} - \dots \right)  \right] \\
	&= - \ln \left[ \upint \!\! \d s  \cE{ \delta_\vec{x} }{s} \right] - \ln \left[  1 - \beta \frac{ \upint \!\! \d s \cE{ \delta_\vec{x} W}{s}}{ \upint \!\! \d s  \cE{ \delta_\vec{x} }{s}} + \frac{\beta^2}{2} \frac{ \upint \!\! \d s \cE{ \delta_\vec{x} W^2 }{s}}{ \upint \!\! \d s  \cE{ \delta_\vec{x} }{s}} - \dots   \right] \, .
\end{align}
We now do the same for the logarithm, $-\ln(1-x)=x+\tfrac{x^2}{2}+\tfrac{x^3}{3}+\dots$,
\begin{align}
	& \beta \Delta G(\vec{x})	+ \ln \left[ \upint \!\! \d s  \cE{ \delta_\vec{x} }{s} \right] = \left( \beta \frac{ \upint \!\! \d s \cE{ \delta_\vec{x} W}{s}}{ \upint \!\! \d s  \cE{ \delta_\vec{x} }{s}} - \frac{\beta^2}{2} \frac{ \upint \!\! \d s \cE{ \delta_\vec{x} W^2 }{s}}{ \upint \!\! \d s  \cE{ \delta_\vec{x} }{s}} + \dots \right) \nonumber \\
	\quad\quad &+ \frac{1}{2} \left( \beta \frac{ \upint \!\! \d s \cE{ \delta_\vec{x} W}{s}}{ \upint \!\! \d s  \cE{ \delta_\vec{x} }{s}} - \frac{\beta^2}{2} \frac{ \upint \!\! \d s \cE{ \delta_\vec{x} W^2 }{s}}{ \upint \!\! \d s  \cE{ \delta_\vec{x} }{s}} + \dots \right)^2 + \dots \\
	&= \beta \,  \frac{ \upint \!\! \d s \cE{ \delta_\vec{x} W}{s}}{ \upint \!\! \d s  \cE{ \delta_\vec{x} }{s}} - \frac{\beta^2}{2} \left[\frac{ \upint \!\! \d s \cE{ \delta_\vec{x} W^2 }{s}}{ \upint \!\! \d s  \cE{ \delta_\vec{x} }{s}} + \left( \frac{ \upint \!\! \d s \cE{ \delta_\vec{x} W}{s}}{ \upint \!\! \d s  \cE{ \delta_\vec{x} }{s}} \right)^2 \right] + \dots \\
	&= \beta \cE{W}{\vec{x}} - \frac{\beta^2}{2} \left[ \cE{W^2}{\vec{x}} + \cE{W}{\vec{x}}^2 \right] + \dots \\
	&= \beta \cE{W}{\vec{x}} - \frac{\beta^2}{2} \cE{\delta W^2}{\vec{x}} + \dots
\end{align}
where 
\begin{align}
	\label{eq:conditinal_biased_average}
	\cE{W}{\vec{x}} &\coloneqq \frac{ \upint \!\! \d s \cE{ \delta_\vec{x} W}{s}  }{ \upint \!\! \d s^\prime \cE{ \delta_\vec{x}}{s^\prime} }
\end{align}
and
\begin{align}
	\delta W(\vec{x})= W -\cE{W}{\vec{x}} \, .
\end{align}

\subparagraph*{Friction matrix estimators}

Starting from the expression of the main text,
\begin{equation}
	\label{eq:friction_collective_variables}
	\itGamma_{ii}(\vec{x}) \approx \frac{\kb T}{m v^2} \frac{\left< \dot{W}_\text{diss} \middle| \vec{x} \right>}{ \cE{\dot{x}_i^2 }{\vec{x}} } \; .
\end{equation}
we explore the alternative of measuring the friction by treating $\dot{W}_\text{diss}$ as a function of $\dot{\vec{x}}$,
\begin{equation}
	\dot{W}_\text{diss} = \beta m v^2 \, \vec{\dot{x}}^\T \matrix{\itGamma} \vec{\dot{x}} + \delta\dot{W}
\end{equation}
and then eliminate the influence of the residual random forces $\delta\dot{W}$ via averaging in the spirit of a linear fit of the parameters $\matrix{\itGamma}$ to this multi-dimensional function. That is to find the $\frac{N(N+1)}{2}$ parameters $\itGamma_{i j}^\text{fit}=\itGamma_{ji}^\text{fit}$, which minimize the $\chi^2$ error to the data,
\begin{align}
	\chi^2 = \left< \left[ \dot{W}_\text{diss} - \beta m v^2\left( \sum_{i} \itGamma_{ii} \dot{x}_i^2 + 2 \sum_{i>j} \itGamma_{ij} \dot{x}_i  \dot{x}_j + \delta\dot{W} \right) \right]^2 \right> \; .
\end{align}
In the special diagonal case we find that the approximate solution
\begin{align}
	\label{eq:friction_collective_variables_fit}
	\itGamma_{ii}^\text{fit}(\vec{x}) \approx \frac{\kb T}{m v^2} \frac{  \cE{ \dot{x}_i^2 \dot{W}_\text{diss} }{\vec{x}}}{ \cE{\dot{x}_i^4 }{\vec{x}} } 
\end{align}
is most instructive to compare to Eq.~\eqref{eq:friction_collective_variables}, though an exact minimization of $\chi^2$ is, of course, also possible.
As a note of caution, we would like to stress that this exploratory ansatz is still lacking a precise theoretical foundation and that it does by no means provide an exact equation of the unbiased friction
matrix.


\section*{Supplementary Figures}


\renewcommand{\thefigure}{S1}
\begin{figure}[ht!]
	\begin{centering}	
		\includegraphics[width=0.9\textwidth]{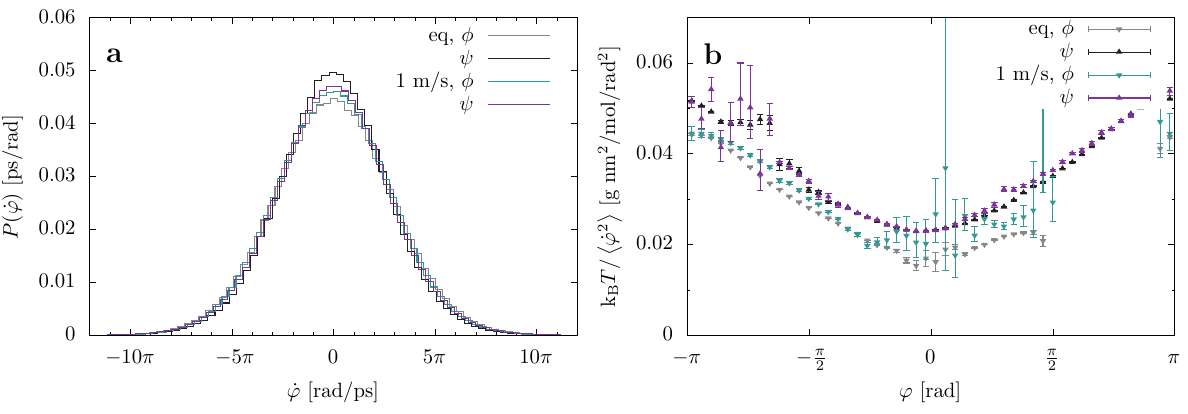}
		\caption{Effective mass of AlaD. (a) Distribution of velocities comparing unbiased trajectory with pulling simulations at $v=1$~m/s. (b) Resultant mass estimates, comparing unbiased case with reweighted estimate of $m_\phi$ and $m_\psi$. }
		\label{fig:alaD_mass}
	\end{centering}
\end{figure}

\renewcommand{\thefigure}{S2}
\begin{figure}[ht!]
	\begin{centering}	
		\includegraphics[width=0.6\textwidth]{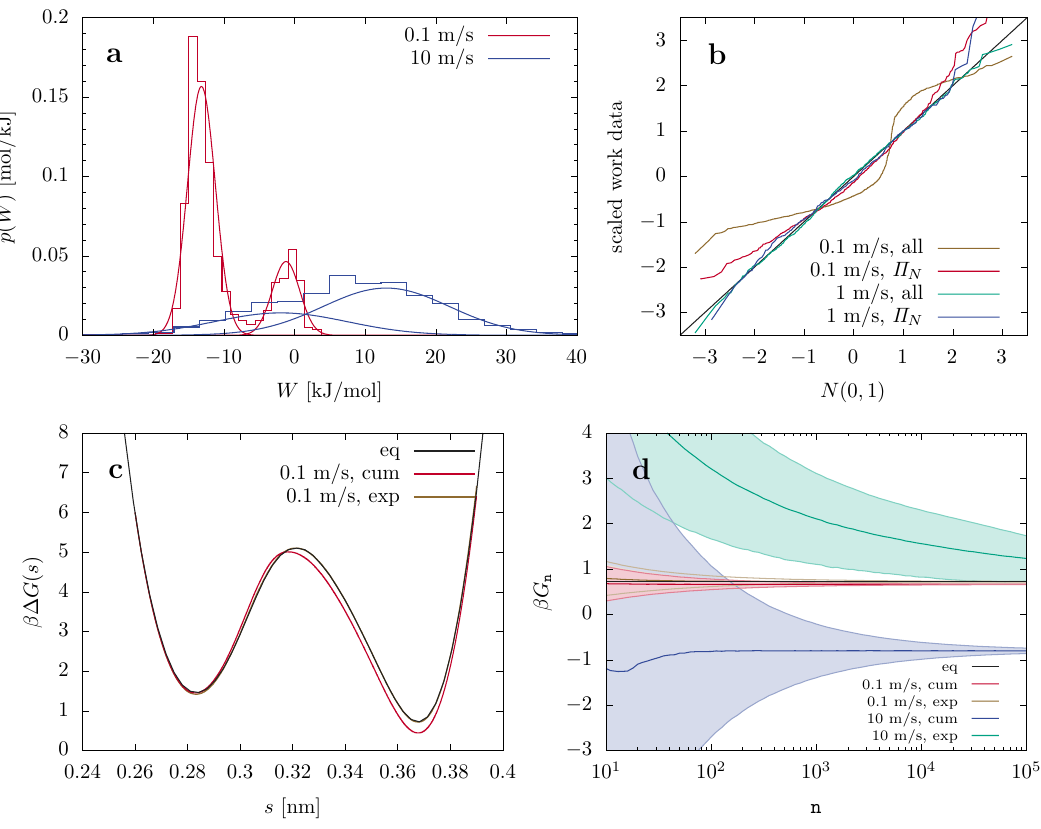}
		\caption{Convergence of AlaD pulling simulations. (a) Work profile at $s=0.368$~nm ($\beta$) comparing $v=0.1$~m/s and $v=1$~m/s. (b) Same data, but shown as normal probability plot, revealing non-Gaussian shapes. (c) Free energy surface estimate pulling at $v=0.1$~m/s, where the cumulant estimate only evaluates path $\itPi_N$. (d) Convergence estimation for the velocities $v=0.1$~m/s and $v=1$~m/s, comparing the asymptotically unbiased but slowly converging exponential average with the bias of the cumulant approximation due the non-Gaussian distribution. Shaded areas denote the bootstrap error estimate.}
		\label{fig:alaD_convergence}
	\end{centering}
\end{figure}

\renewcommand{\thefigure}{S3}
\begin{figure}[h!]
	\begin{centering}	
		\includegraphics[width=0.6\textwidth]{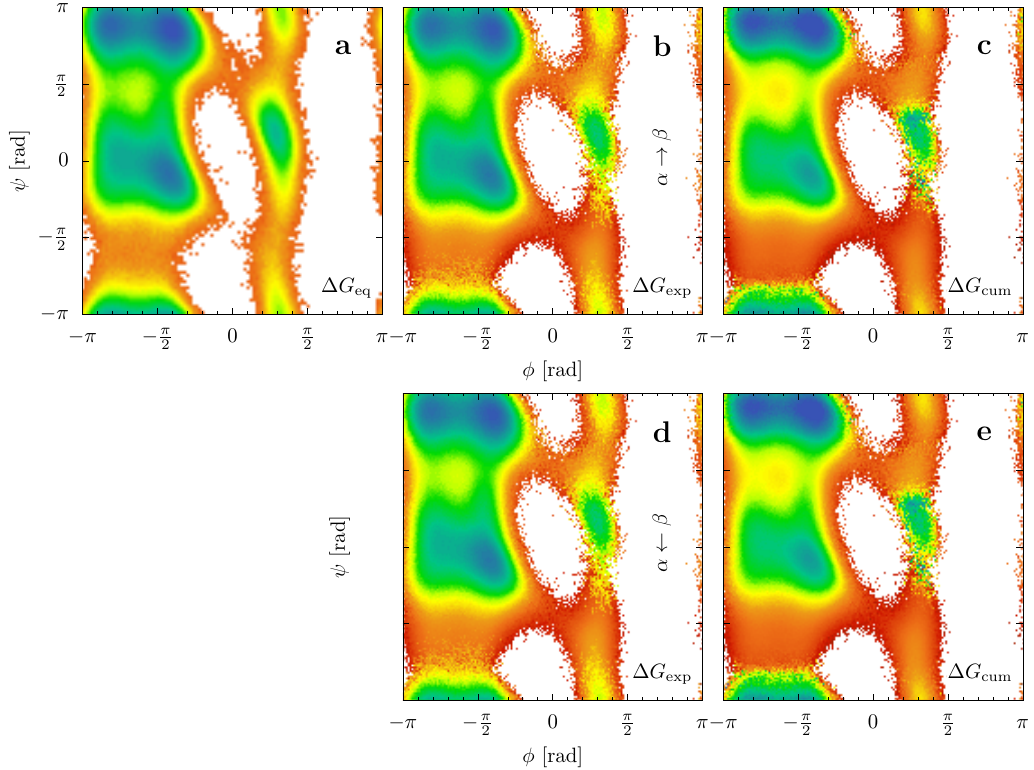}
		\caption{Free energy landscape of AlaD. (a) Free energy landscape estimated from unbiased MD simulation compared to $\alpha \rightarrow \beta$ pulling simulations (b,c), as well as the reverse direction (d,e). (b,d) Estimate using the exponential average, Eq.~(4a). (c,e) Estimate using the cumulant approximation, Eq.~(4b).}
		\label{fig:alaD_both_directions}
	\end{centering}
\end{figure}

\renewcommand{\thefigure}{S4}
\begin{figure}[ht!]
	\begin{centering}	
		\includegraphics[width=0.6\textwidth]{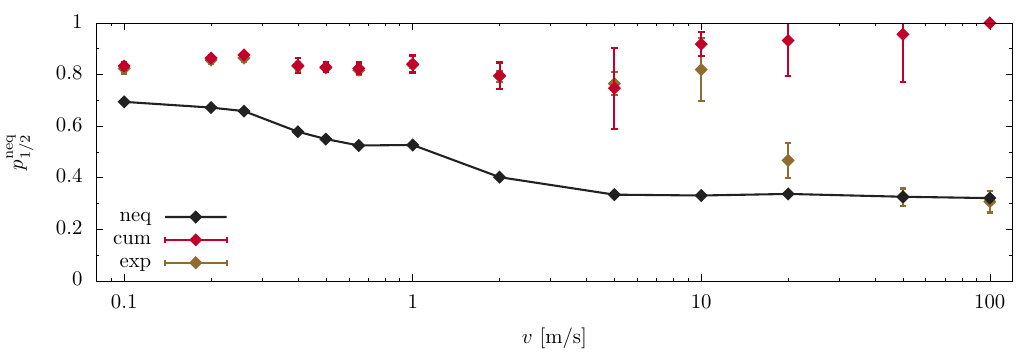}
		\caption{Fraction of pulling trajectories going through the native path. Lines to guide the eye.}
		\label{fig:alaD_path_ratio}
	\end{centering}
\end{figure}

\renewcommand{\thefigure}{S5}
\begin{figure}
	\begin{centering}	
		\includegraphics[width=0.98\textwidth]{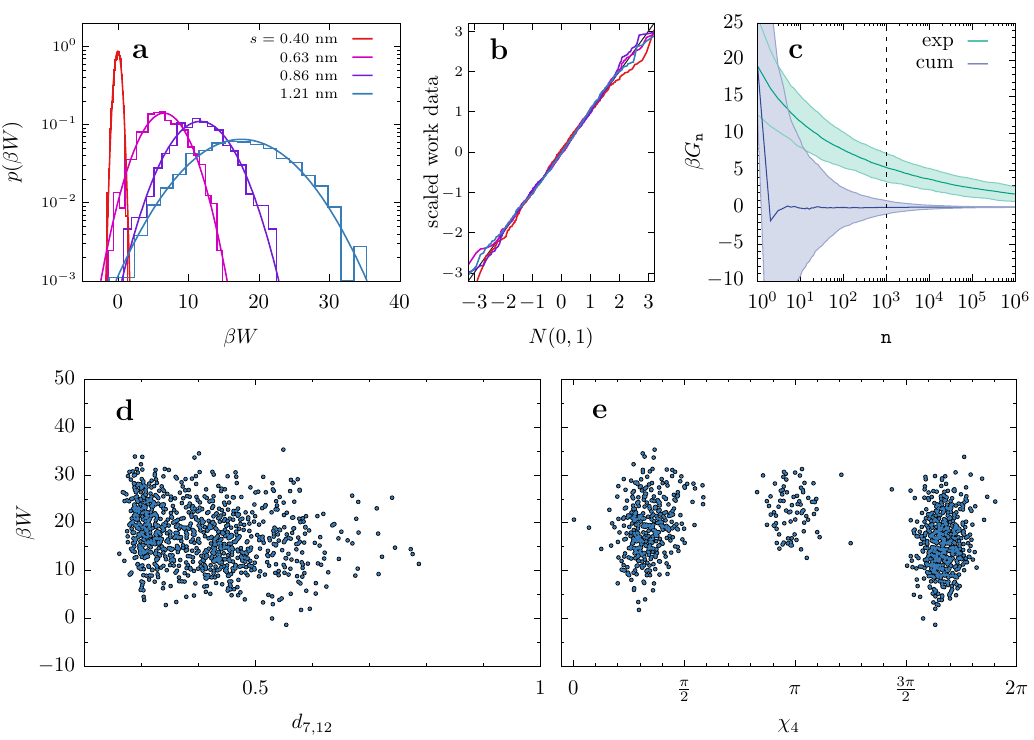}
		\caption{Convergence estimation of T4l results. (a) Work histogram for different locations in $s$. (b) The same data, visualized by a normal probability plot, showing all data standardized and sorted in ascending order and against the quantiles of a normal distribution. (c) Convergence of free energy estimate at $s=1.21$~nm, showing running average statistics as a function of sample size, comparing exponential average (exp) with cumulant approximation (cum). Dashed line represents average of 1000 repetitions of the running average, showing the expected bias. Colored areas represent the corresponding bootstrap error of the free energy estimates. (d) Work values stratified using $d_{7,12}$ and (e) $\chi_4$, respectively.}
		\label{fig:SI_t4l_convergence}
	\end{centering}
\end{figure}

\renewcommand{\thefigure}{S6}
\begin{figure}
	\begin{centering}	
		\includegraphics[width=0.98\textwidth]{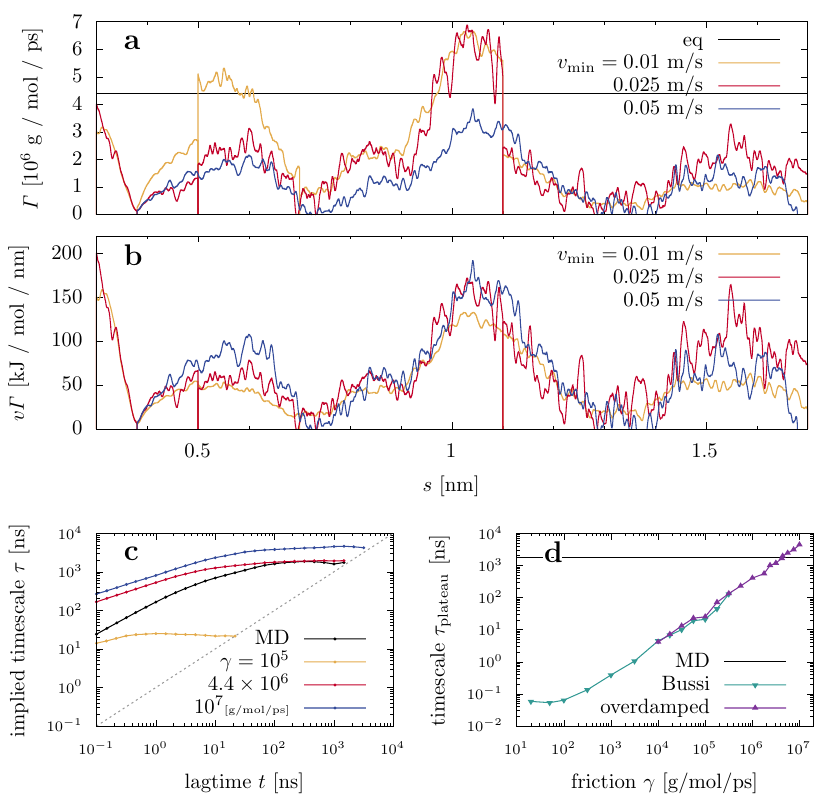}
		\caption{Friction modeling of T4L. (a) Friction profile from TMD simulation, pulling with $v_\text{min} = 0.025$~m/s from 0.5 to 1.1~nm, rest with $v = 0.05$~m/s, as well as with $v_\text{min} = 0.05$~m/s everywhere, compared to a constant friction coefficient resulting in the correct \textbf{1} $\leftrightarrow$ \textbf{4} timescale in unbiased Langevin simulations. (b) Friction multiplied with constraint velocity. (c) Implied timescales as a function of lag-time using the slowest time scale of the four state model in $s$, comparing unbiased MD reference with Langevin simulations with different constant friction $\gamma$. (d) The resulting \textbf{1} $\leftrightarrow$ \textbf{4} timescale as a function of $\gamma$, read from the implied timescale plateau. The black line indicates the timescale matching the unbiased MD, where the corresponding $\gamma$ is used as a reference. Note that for large friction, numerical integration of a non-overdamped Langevin equation by the Bussi-Parinello integrator (OBABO) became instable. We therefore used an overdamped first-order Runge-Kutta integrator.}
		\label{fig:SI_t4l_friction}
	\end{centering}
\end{figure}

\end{document}